# Is it Cake or is it AI? A Systematic Review of Human Uncertainty in Distinguishing Generative Artificial Intelligence Content

Mark Louie F. Ramos

Department of Health Policy and Administration, The Pennsylvania State University, University Park, PA

mlr6219@psu.edu

**Abstract**

This systematic review synthesized empirical evidence on human ability to distinguish generative artificial intelligence content from human-produced content across text, image, and voice modalities. A structured search of Scopus identified 22,541 records from 2025-2026, of which 1200 were screened and 30 studies were included. Across these studies, human detection accuracy varied widely but generally clustered around chance performance. Overall, the literature shows that humans are generally unreliable detectors of gen AI content, raising broader questions about whether the ability to tell should matter for how we evaluate or trust content.

## Introduction

Since the rise in popularity of Large Language Model (LLM) chatbots like ChatGPT or Google Gemini, also commonly referred to as artificial intelligence or AI, one of the most controversial and empirically studied topics in how people interact with these new technologies is the uncertainty people experience when judging whether a piece of content was produced by generative artificial intelligence (gen AI) [1–4]. This has only seemed to have accelerated, with at least 16 studies published in the first 3 months of 2026 alone [5–19]. Diverse concerns motivate these studies, ranging from worries about the large-scale dissemination of potentially malicious AI-generated content such as social-engineering and disinformation mechanisms in online environments [7,20] to concerns about authenticity in personal application statements [12,14] and academic integrity in scientific manuscripts and essays [13,15]. Despite this proliferation of empirical work, there is yet to be a synthesis of how accurately humans can identify AI-generated content across modalities or contexts. To address this gap, this systematic review synthesizes current evidence on human ability to recognize different kinds of gen AI content.

## Methods

### Search Strategy

This review followed the Preferred Reporting Items for Systematic Reviews and Meta-Analyses (PRISMA) guidelines for study identification, screening, eligibility assessment, and inclusion. The overall study selection process is summarized in Figure 1.

A literature search was conducted in the electronic database Scopus, which was used as the primary source for indexed publications because it offers broad multidisciplinary coverage across fields such as computer science, psychology, linguistics, and human–computer interaction. Moreover, the database supports complex Boolean queries with wildcards, enabling better, context-specific retrieval of studies involving generative AI systems, human evaluators, and detection or discrimination tasks. The search was restricted to 2025–2026 to ensure the review reflects the most current evidence on human detection of generative AI content. The keywords used during the search included terms related to AI-generated media, human detection, and evaluation tasks. The final Boolean query was: (TITLE-ABS-KEY("AI generat*" OR "machine generat*" OR "synthetic text" OR "synthetic media" OR "large language model" OR "LLM" OR ChatGPT OR GPT OR "generative AI" OR "neural text generator")) AND (TITLE-ABS-KEY(detect* OR identify* OR distinguish* OR recogniz* OR "classification task" OR "human evaluation" OR differentiat* OR discriminat*)) AND (TITLE-ABS-KEY(human OR particip* OR user OR evaluator OR judge OR rater OR annotat* OR "lay person" OR "expert reviewer"))

## Study Eligibility Criteria

The review included published papers involving human participants who were asked to detect or distinguish AI-generated versus human-generated content across various modalities (text, image, voice). Eligible studies were required to report quantitative performance metrics from which a retrievable proportion of correct human classifications and the corresponding number of human evaluators could be obtained. Studies were not required to designate accuracy as the primary outcome, provided that sufficient data were available to compute a success proportion and sample size. Only studies published in English during 2025–2026 were considered. Studies were excluded if they were review articles, conceptual or theoretical papers, conference abstracts without results, or empirical studies that used AI-generated stimuli but did not measure human detection accuracy. Papers with incomplete information or insufficient methodological detail were also excluded.

## Data Extraction

The initial search yielded 22,541 records, ordered by relevance. Titles and abstracts were screened sequentially until 1,200 records were reviewed, at which point no eligible studies had been identified in the last 100 records, suggesting saturation. A total of 44 studies met the preliminary inclusion criteria and were retrieved for full-text evaluation. Following further evaluation, 14 studies were excluded for reasons such as lack of a human detection task (3), absence of extractable accuracy data (4), or general methodological ineligibility (7). A total of 30 studies were included in the final synthesis.

Data was extracted and summarized in Table 1. Extracted variables include publication details, modality of AI-generated content, population characteristics, sample size, and reported results involving human accuracy in detecting gen AI content. Where not explicitly provided, we used relevant quantities from the study data and results to obtain the average proportion of correct classifications and the corresponding number of human evaluators.

*Table 1: Included Studies*

| Authors; Publication, Year | Content | Population | Sample Size | Relevant findings |
|---|---|---|---|---|
| Barrington S et al., Sci Rep, 2025 [21] | Voice | Adult listeners | 294 | This study reveals that, generally speaking, people are poorly equipped to identify AI-generated voice clones. |
| Lavan N et al., PLoS One, 2025 [22] | Voice | Adult listeners (UK-based, native English speakers) | 50 | We show that, under certain conditions, it is not possible for human listeners to accurately discriminate between AI-generated |
| | | | | voices and genuine recordings of human voices. |
| Yoon D, Oh GE, Kent R., Lingua, 2026 [5] | Voice | Adult native Korean listeners | 36 | On average, participants correctly identified the voice type in 75.4 % of trials. |
| Froehnel K et al., HICSS, 2025 [20] | Restaurant reviews | Online consumers | 151 | We found that humans cannot reliably distinguish between genuine and AI-generated fake reviews (accuracy = 53.2%). |
| Kober S.E. et al., Comput. Hum. Behav., 2026 [6] | Conversational text | Female adult participants | 21 | Participants who were unaware of the true identity of their communication partner were unable to correctly identify when they had interacted with an AI and when they had interacted with a human experimenter |
| Saucier C.J. et al., Comput. Hum. Behav., 2026 [7] | Social media posts | Undergraduate students in U.S. and U.S. adults | 565 | Participants performed at essentially chance levels when judging stimuli accounts as bots versus humans. |
| Szabó G. et al., Soc. Sci. Humanit. Open, 2026 [8] | Narrative texts in Hungarian | Adult Hungarian-speaking readers | 576 | On average, participants correctly identified the author of the texts in 66% of the cases. |
| Tuomi A et al., ICTT, 2025 [23] | Online restaurant review texts | Adult human participants | 80 | Out of the overall 800 evaluations our participants made, only around half were correct |
| Holyoak KJ., Journal of Creativity, 2026 [9] | Poetry | Expert poets (poetry critics) | 2 | Both expert poets correctly identified the AI-generated poem. |
| Zanardo M et al., Eur Radiol Exp, 2026 [10] | Structured lumbar spine MRI radiology reports | Medical professionals (radiologist, residents, GP, orthopedic surgeon) | 5 | Some AI-generated reports were indistinguishable from human ones, particularly for non-specialized readers |
| Linde P et al., npj Digit Med, 2026 [11] | Multiple-choice questions (MCQs) for medical imaging education | Medical students and physicians in imaging-related specialties | 128 | Participants did not identify item origin above chance. |
| Wachholz F et al., BMC Res Notes, 2025 [24] | Exercise training plans (12-week half-marathon plan) | Recreational athletes and experienced coaches | 6 | The quality of the output from large language models has now reached a level where even professional coaches are often unable to distinguish whether a training plan was AI-generated or created by a human expert. |
| Karakash WJ et al., Global Spine J, 2026 [12] | Fellowship personal statements for spine surgery applications | Spine surgery faculty and fellows | 8 | Reviewers could not distinguish AI vs human authorship. |
| Zhu T et al., Findings ACL, 2025 [25] | Varied text | English-speaking adult MTurk workers | 600 | Blind tests show humans cannot distinguish AI vs human text. |
| Franke F. et al., Internet Interventions, 2025 [26] | Psychological advice responding to newspaper mental-health questions | Licensed mental health clinicians (psychologists and psychotherapists) | 45 | Participants could not distinguish between AI- and expert-authored advice (p = .27). |
| Fiedler A., Döpke J., Int. Rev. Econ. Educ., 2025 [27] | Academic text excerpts resembling German theses and academic writing | University lecturers with teaching experience | 63 | The results show that both human evaluators and AI detectors correctly identified AI-generated texts only slightly better than chance, with humans achieving a recognition rate of 57 % for AI texts and 64 % for human-generated texts. |
| Stadler RD et al., Arthroscopy, 2025 [28] | Scientific research abstracts in shoulder and elbow surgery that mimic published abstracts | Experienced peer reviewers in shoulder and elbow orthopaedic surgery | 8 | Experienced reviewers faced difficulties in distinguishing between human and AI-generated research content within shoulder and elbow surgery. |
| Cardia F et al., LNCS (EC-TEL), 2025 [29] | Machine-generated student questions based on | University instructors | 7 | Results show that instructors struggle to differentiate between the two sets of questions, with accuracy close to random chance. |
| | university video lectures | | | |
| Šindlerová J et al., Springer Proc. Complexity, 2026 [13] | Student essays written to resemble 17-year-old high school students | Master's-level pedagogy (teacher-education) students | 60 | The average classification accuracy of 49.9% suggests that participants struggled to distinguish between machine-generated and human-written essays, performing essentially at the level of random guessing. |
| Vaccaro MJ et al., J Am Coll Surg, 2026 [14] | Medical school personal statements | Medical school application readers (admissions committee members) | 17 | Readers identified authorship with 56% accuracy. |
| Goodman MA et al., J Phys Ther Educ, 2025 [30] | Personal statements modeled after physical therapist education program applications | Raters of applicants to a doctor of physical therapy program | 2 | Human raters showed very high interrater reliability (κ = 0.92) and accuracy of 97% and 99% in identifying AI-generated statements. |
| Alkhofi A., Forum Linguist Stud., 2025 [31] | Translations (English→Arabic) | University ESL instructors | 20 | Converting the log-odds of −0.948 to a probability using the exp function indicates that instructors, on average, have a 28% chance of correctly identifying machine translations, which is significantly below the 50% expected as a result of random guessing. |
| Jain A., J Cranio-Maxillofac Surg, 2026 [15] | Full-length academic manuscripts (~2500 words) in oral and maxillofacial surgery | Experienced oral and maxillofacial surgeons | 20 | Reviewers correctly identified manuscript authorship only 54 % of the time |
| Madleňák M. et al., Transp. Res. Procedia, 2026 [16] | Phishing-related content across multiple media | Security professionals involved in critical infrastructure, transport security, crisis management, and security management | 21 | The results revealed that the respondents were able to correctly detect a total of 145 content cases, representing a rounded 57% of the total number of responses. |
| Langer A et al., Technology in Society, 2025 [17] | Images of objects, art, and faces | Children aged 6–10 years | 37 | On average, children performed at or below chance in discriminating human-created from AI-generated content across modalities, and their overall accuracy was significantly below that of adults. |
| Velásquez-Salamanca D et al., J Imaging, 2025 [32] | Human portraits, landscapes, everyday scenes, detailed objects | Adult participants including visual professionals and non-professionals | 161 | We conclude that individuals are generally unable to accurately determine the source of an image, which in turn affects their assessment of its credibility. |
| Högemann M et al., Front Artif Intell, 2025 [33] | Photorealistic images of landscapes, architecture, and interiors | German-speaking adults | 104 | A quantitative analysis revealed that participants correctly identified AI-generated images in 63.7% of cases overall and notably in only 29% of cases<br>when FLUX.1-dev was used. The hierarchical model estimated lower odds of<br>correct detection with increasing age, while education, gender, AI-tool use,<br>media work, and editing experience showed no significant effects. |
| Wang Z & Jin Y., Proc. CHCHI, 2025 [34] | Abstract art wallpapers | Adult users recruited from a high-tech company in China | 34 | Moreover, most participants cannot distinguish AI-generated art from human-made designs. |
| Sarno DM et al., Acta Psychologica, 2026 [18] | Human face images | Adult US, native English speakers | 166 | Participants were so poor at the image classification task that their identification of the AI-generated faces dipped below chance performance. |
| Chow JK et al., J Exp Psychol Gen, 2026 [19] | Human face images | Adult human participants | 250 | We show that some individuals are consistently better at discriminating real from AI-generated faces. |

*Figure 1: PRISMA flow diagram*

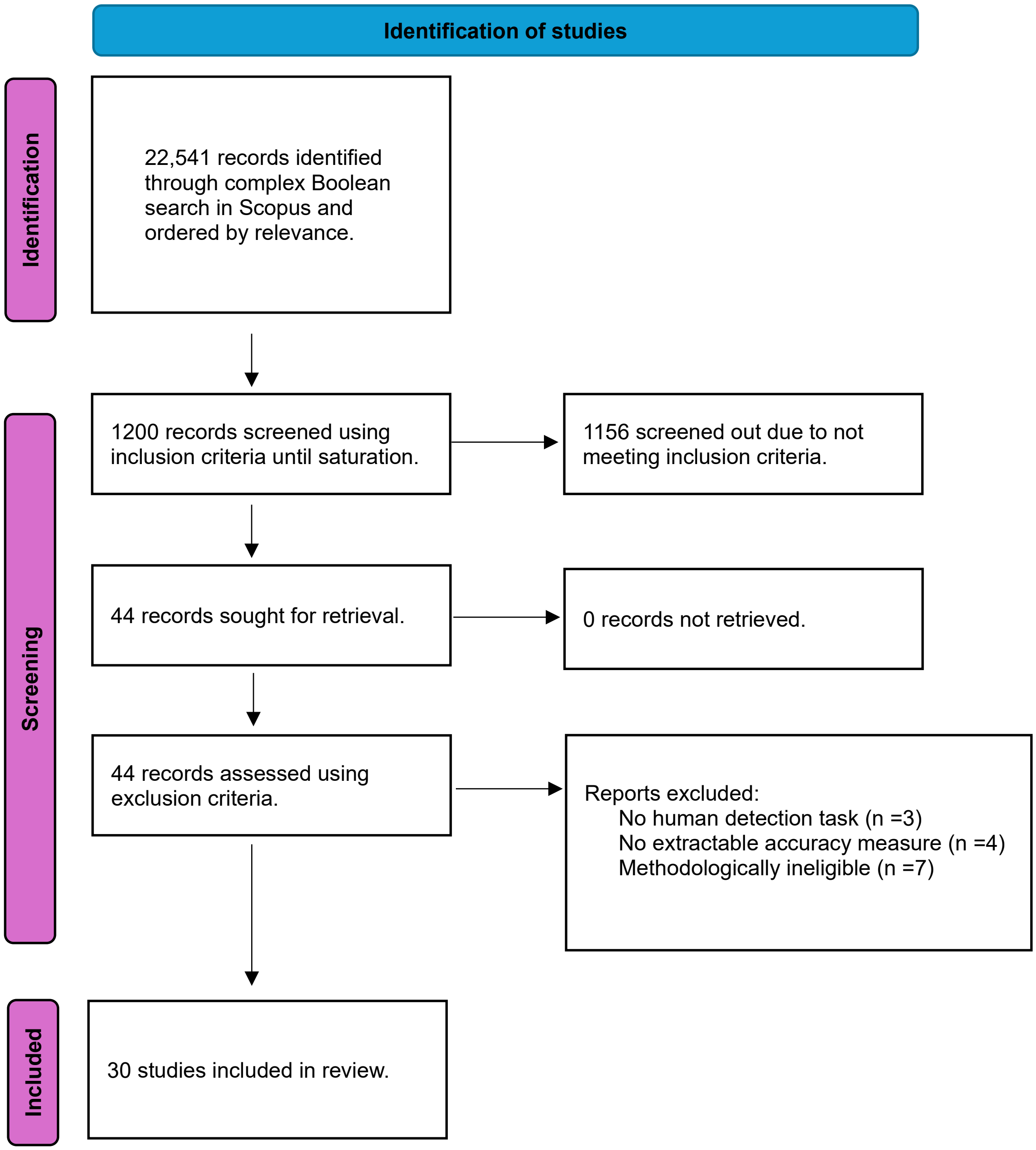

**Results**

The studies included in this review were highly heterogeneous in design, modality, participant populations, and outcome definitions. Because of this substantial methodological and conceptual diversity, computing pooled effect sizes was not attempted. Instead, the evidence was summarized descriptively, reporting the total number of human participants across studies and the observed range of accuracy estimates. For the accuracy estimates, most studies reported pooled proportions across all classification attempts, without providing accuracy values at the level of individual human raters. Because participant-level accuracy could not be reconstructed from that information, the accuracy estimates presented necessarily assume a strong degree of regularity across participants within studies.

Out of 30 studies reviewed, 17 [6,7,11–13,17,18,20,21,23–27,29,31,34] with a total sample size of 2285 individuals explicitly reported lack of confidence that human ability to detect AI generated content was better than chance. The accuracy rates reported by these studies ranged from 28% [31] to 67% [24]. In contrast, six studies [5,8,9,19,22,30] with total sample size of 916 explicitly indicated the contrary, with accuracy rates that ranged from 61% [19] to 100% [9]. The remaining studies, total sample size 336, did not explicitly state a directional conclusion regarding human uncertainty, but they reported accuracy estimates ranging from 54% [15] to 78% [10]. In comparison, Figure 2 shows a forest plot summarizing estimated human accuracy derived from the classification data reported across the reviewed studies. There are four studies [6,7,16,17] that appear twice due to multiple sets of classification data reported in these studies for different settings, thus resulting in 34 estimates. Among these estimates, 8 produced confidence intervals that are statistically above 50% [5,6,8,19,21,22,32,33], while 1 produced a confidence interval that is statistically below 50% [7].

When we categorize the results by gen AI content type (Figure 3), we see that 3 out of the 8 estimates with confidence intervals above 50% were on voice content. Furthermore, these 3 are the only studies included in the review where voice is the content for classification. Of the remaining 5, 3 were on image content (out of 7 results on image classification) and 2 were on text content (out of 24 results on text classification).

We further inspect the results on text classification where we distinguish between media text content (posts, conversations, etc.) and technical text content (reports, application statements, manuscripts). Figure 4 shows that the two studies where the chance of human classification was higher than 50% were both media-type content, specifically narrative texts in Hungarian evaluated by adult Hungarian-speaking readers [8] and conversational text evaluated by female adults where they were informed beforehand that their conversation partner may be an AI chatbot [6], where accuracy was estimated at 85%. In this same study, for participants who were not notified beforehand, accuracy was 52% [6]. There are two special cases in Figure 4 which we could not classify as media or technical, one was on translations, where the accuracy rate was 28% but only involved 20 raters and the other was on poetry where the rate was 100% but only involved 2 raters.

*Figure 2: Forest plot of estimated average human accuracy in detecting AI-generated content. Each point reflects study-level observed accuracy. Error bars are 95% Wilson Confidence Intervals computed using the number of raters as sample size.*

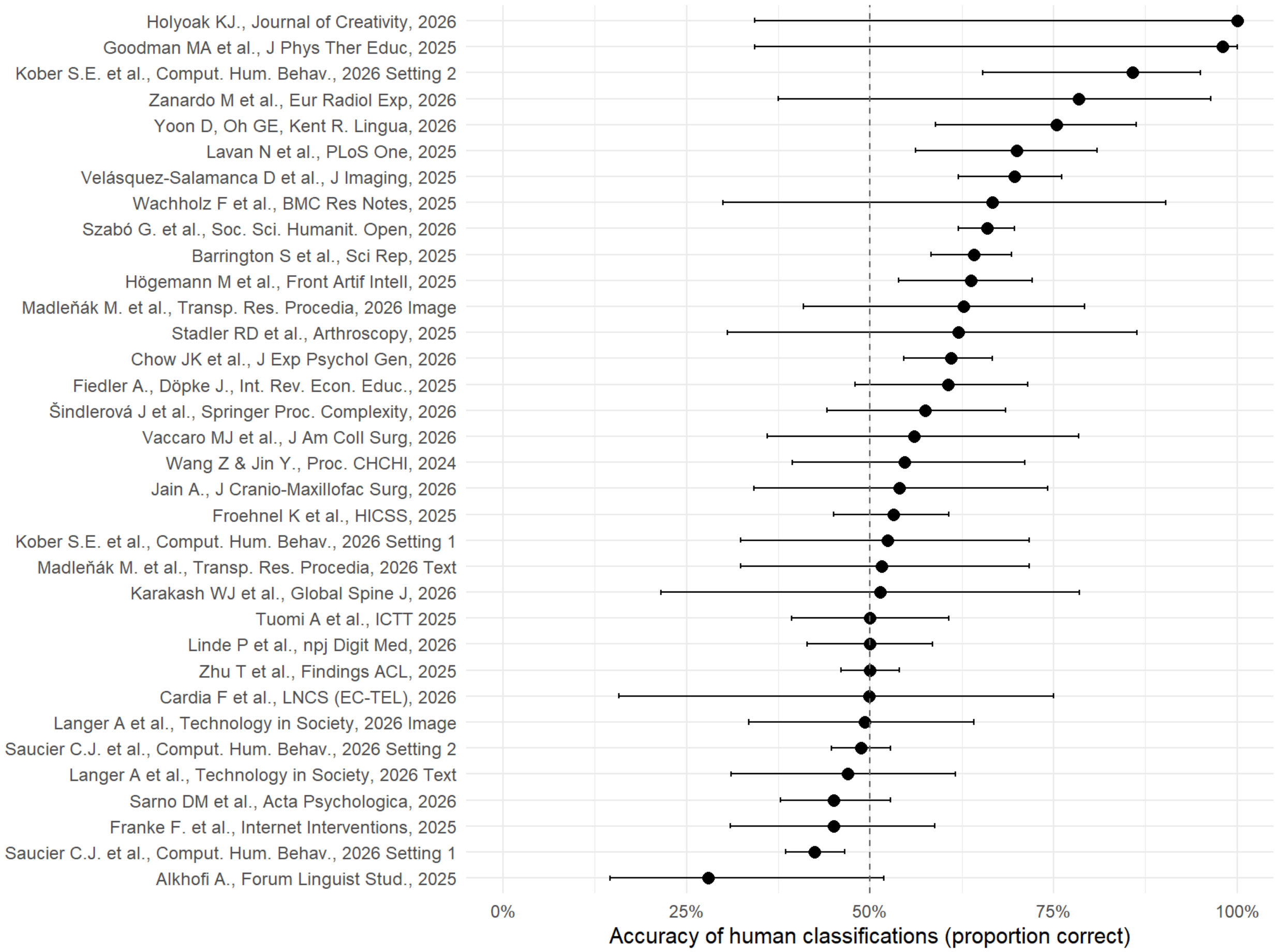

*Figure 3: Accuracy by Gen AI Content Type.*

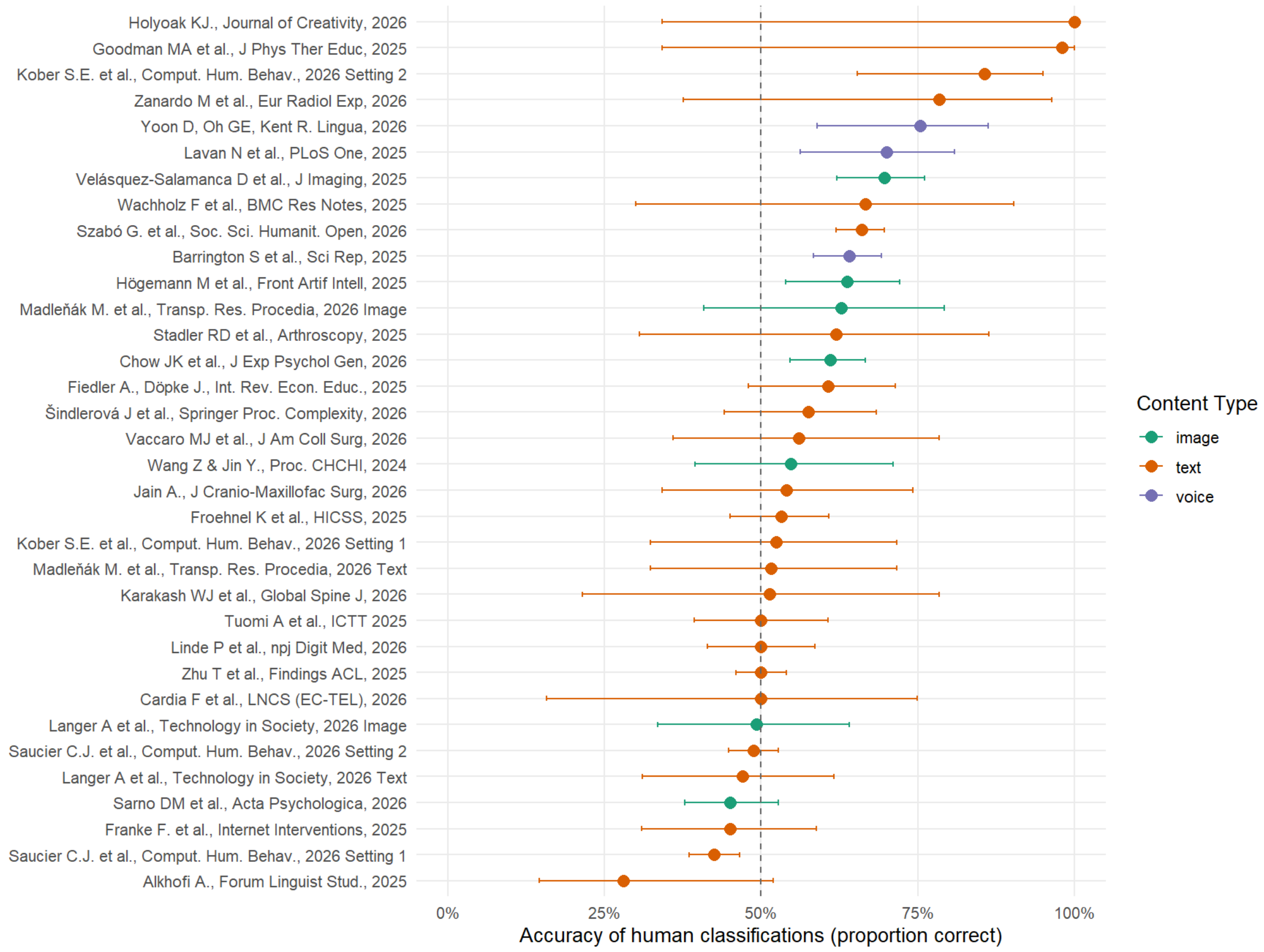

*Figure 4: Accuracy by Text Content Subtype.*

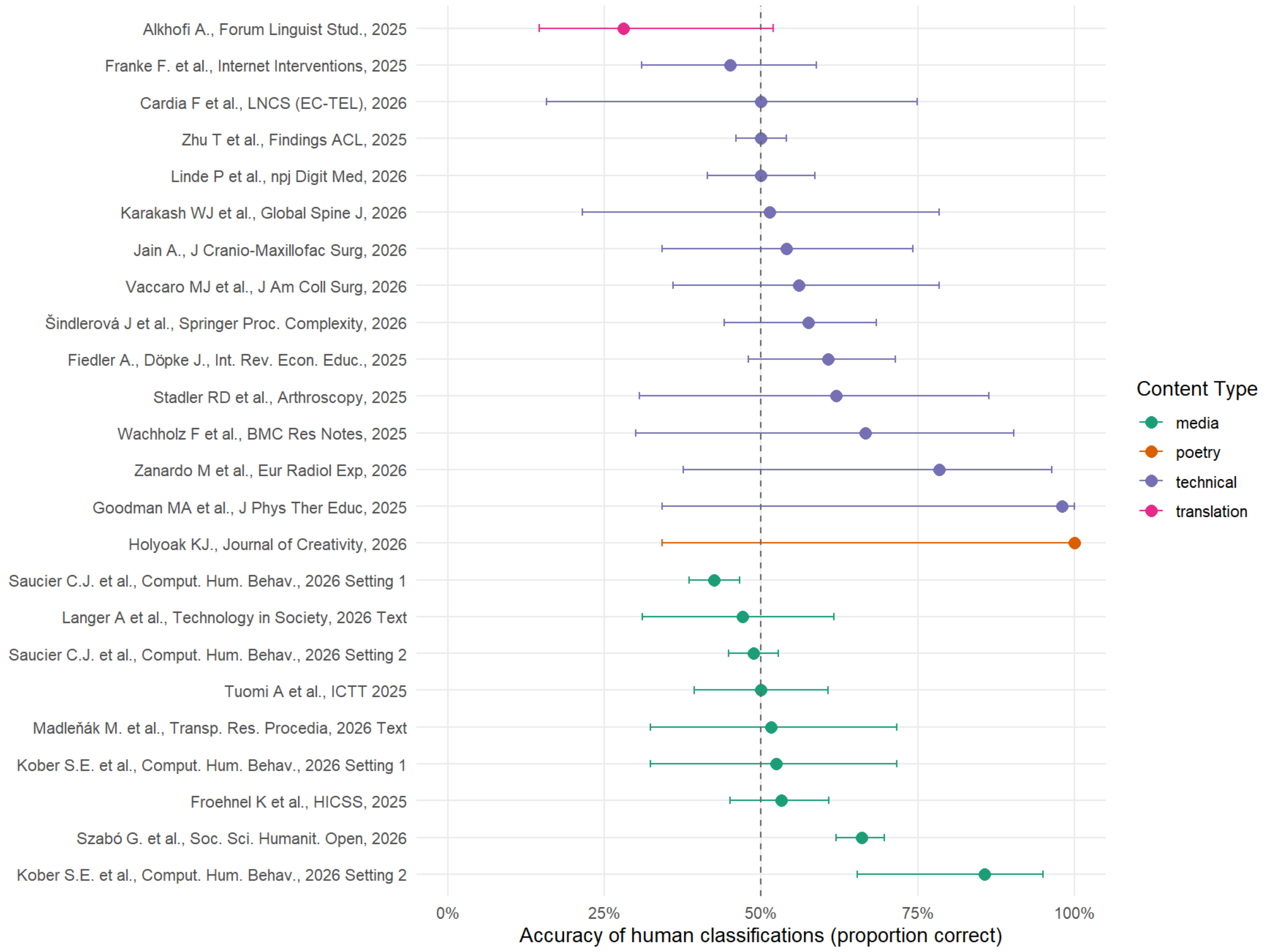

## Discussion

Despite substantial methodological and contextual heterogeneity across the studies reviewed, a consistent pattern observed is that humans, on average and in general, do not distinguish AI-generated content from human-generated content reliably better than chance. This echoes earlier assessments and positions [35–37], suggesting that human accuracy has not meaningfully improved over time or at least has not improved at a pace that keeps up with the increasing realism of gen AI content [38].

Performance was observed in reviewed studies to vary by modality, with more success at identifying AI-generated voice content than images, and more success with images than with text. This is consistent with [39] which discussed how humans may be better at detecting synthetic voices than images or text due to subtle artifacts in timing, pitch, or prosody. For image detection, visual heuristics, such as inconsistent reflections, unnatural textures, or impossible lighting may help identify generative AI content [40], but accuracy rates from reviewed studies on images still mostly did not vary convincingly from chance, also consistent with previous studies [41].

In line with motivations for conducting these studies, an important subgroup of the reviewed literature focused on text content which has typically served as basis for evaluators' judgments about the author's expertise or competence. These include personal statements for applications [12,14,30] and academic manuscripts or essays [13,15,27,28]. Accuracy across these studies weighted by sample size is 58%. In addition, evaluators oftentimes perceived gen AI generated documents not only as human written, but of better quality [12,14,15]. It is important to point out that the reviewed studies' prevailing discussion of their results is not to call on improving accuracy, but to challenge the prevailing use of how well these texts are written as a meaningful indicator of quality. Personal statements should only be valued insofar as they contain verifiable, objective information about a candidate's experiences, decisions, and conduct. The candidate's rhetorical sophistication should not be construed, explicitly or implicitly, as evidence of their suitability. Long before generative AI, applicants with access to professional editing or commercial statement-writing services were already advantaged by the same evaluative bias toward rhetorical refinement [12]. By lowering the cost of high-quality writing assistance, gen AI may in fact be partially leveling the playing field for applicants with fewer financial resources or lower linguistic ability, especially for those whom English is not a first language. Similar reasoning can be applied to scientific writing. The point of polished language in research manuscripts has always been merely to improve clarity. Scrutiny of the reasonableness of methods and the veracity of results is invariant to how the text was produced. As discussed in [42], gen AI may "serve as an equalizer" for researchers who struggle with academic English.

Naturally, this dynamic is different when generative AI is used with the intent to deceive, such as in fabricated customer reviews [23] or coordinated social-media manipulation campaigns [7]. In these contexts, the concern is not evaluative fairness but the risk that highly convincing and abundant AI-generated content can amplify misleading or harmful messages. The results of this review underscore the need for readers to adopt a more cautious, discerning stance toward online information, resisting the impulse to treat unverified claims as credible simply because they appear to be widely repeated. Misinformation and manipulative amplification predate gen AI [43,44], but the scale enabled by these tools makes the need for careful, educated, and verification-oriented reading practices more urgent and obvious. In this sense, the challenge posed by gen AI content may ultimately prompt healthier norms of online discernment.

## Limitations

This review has limitations that should be considered when interpreting its findings. First, the search was conducted exclusively in Scopus, which, although broad in multidisciplinary coverage, may not index all relevant studies, particularly those in fast-moving computer science venues or preprint repositories. Second, while there were more than 20,000 hits from the initial search, only the first 1,200 ordered by relevance were screened. While the consistency of the findings across the reviewed studies suggests that the conclusions are stable, there may be studies missed that could provide further nuance. Third, substantial methodological heterogeneity across studies limited the ability to compute pooled effect sizes. Accuracy estimates therefore assume regularity across participants within each study, which may not reflect true variability in individual detection ability.

## Conclusions

Considerable human uncertainty in identifying gen AI content is real and is observable across modalities and, at least for text, contexts within the modality. This raises a broader question about what matters going forward: whether people can tell if something is gen AI content or whether the ability to tell should remain relevant to how we evaluate or trust content. Perhaps the better path is becoming more honest about the criteria we use to judge quality and more discerning about what we choose to believe from what we read, regardless of who or what actually produced the words.